\documentclass[twocolumn,prl,aps,amsmath,showpacs,amssymb,nofootinbib]{revtex4}
\usepackage{graphicx}
\usepackage{color}
\usepackage{amsmath,amssymb}
\usepackage{array}

\newcommand{\TeV}{\, \text{TeV}}
\newcommand{\GeV}{\, \text{GeV}}

\newcommand{\tr}{\mbox{tr}}
\newcommand{\gtwo}{I\kern-.1em I\,}
\newcommand{\beq}{\begin{eqnarray}}
\newcommand{\eeq}{\end{eqnarray}}
\newcommand{\bpm}{\begin{pmatrix}}
\newcommand{\epm}{\end{pmatrix}}
\newcommand{\cl}{\, \rm C.L.}

\begin{document}
\title{Strong dynamics, minimal flavor and $R_b$} 

\author{Hidenori S. Fukano}
\author{Kimmo Tuominen}
\affiliation{Department of Physics, University of Jyv\"askyl\"a, P.O.Box 35, FIN-40014 Jyv\"askyl\"a, Finland \\
and Helsinki Institute of Physics, P.O.Box 64, FIN-00014 University of Helsinki, Finland}

\begin{abstract}We discuss how models of electroweak symmetry breaking based on strong dynamics lead to observable contributions to the $Z$-boson decay width to $b\bar{b}$ pairs even in the absence of any extended sector responsible for dynamical generation of the masses of the Standard Model matter fields. These contributions are due to composite vector mesons mixing with the Standard Model electroweak gauge fields and lead to stringent constraints on models of this type. Constraints from unitarity of $WW$-scattering are also considered.
\end{abstract}

\maketitle


The Standard Model (SM) of elementary particle interactions is believed to be an incomplete theory due to its inability to explain the origin of the observed mass patterns of the matter fields, the number of matter generations and why there is excess of matter over antimatter. One possible paradigm beyond the Standard Model (BSM) is  to apply strong coupling gauge theory dynamics. 
In Technicolor (TC) \cite{Weinberg:1975gm} 
, the electroweak symmetry breaking is due to the condensation of new matter fields, the technifermions. The vintage TC model  based on the QCD-like gauge theory dynamics is incompatible with the electroweak precision data from the LEP experiments \cite{Peskin:1990zt}, and most of the modern model building within the Technicolor paradigm concentrates on the so called walking Technicolor (WTC) \cite{Holdom:1984sk}. Here the Technicolor coupling constant evolves very slowly due to a nontrivial quasi stable infrared fixed point \cite{Banks:1981nn}. Models of WTC with minimal new particle content can be constructed by considering technifermions to transform under higher representations of the TC gauge group \cite{Sannino:2004qp}. The walking Technicolor scenarios lead also to a light scalar state compatibly with the LHC discovery of a Higgs -like scalar particle \cite{:2012gk,:2012gu}.
Technicolor only explains the mass patterns in the gauge sector of the SM via strong dynamics at the electroweak scale $\Lambda_{\textrm{TC}}\simeq 1$ TeV. To explain various mass patterns of the known matter fields within a TC framework, further dynamical mechanism are needed; a well known example is the extended TC (ETC) \cite{Dimopoulos:1979es}. An alternative to ETC, aimed to explain the large top quark mass and, in particular, the top-bottom mass splitting is the topcolor model and topcolor assisted technicolor model (TC2) \cite{Hill:1991at}.

One of the main experimental constraints on TC/ETC, and also on TC2, arises from the $Z$ boson decay rate to $b\bar{b}$ pairs, more precisely one considers $R_b \equiv \Gamma(Z\to \bar{b}b)/\Gamma(Z\to {\rm had})$ \cite{Chivukula:1994mn,Burdman:1997pf}.
The importance of various contributions to this observable is determined by the relevant energy scale associated with different stages of the underlying dynamics: The effects from ETC gauge bosons are suppressed by the ETC scale $\Lambda_{\textrm {ETC}}\gg \Lambda_{\textrm{TC}}$, and similarly for the effects of the extended gauge interactions due to the topcolor dynamics. However, the effects from extra goldstone bosons due to topcolor, so called top-pions, are governed by the electroweak scale rather than the topcolor scale, and it has been shown that their effect generally is a substantial reduction of $R_b$ relative to the SM prediction and hence this provides stringent constraints on topcolor dynamics \cite{Burdman:1997pf}.

In this letter we analyse how a TC model is already sensitive and subject to similar constraints already without any extension towards the matter sectors of SM. This is so, since any TC model features composite vector and axial vector states in the spectrum which will mix with the SM gauge fields. We will explicate this issue within a generic low energy effective theory corresponding to the symmetry breaking pattern SU(2)$_L\times$SU(2)$_R\rightarrow$SU(2)$_V$ and discuss the consequences for model building and phemomenology. 

The experimental value of $R_b$ \cite{Baak:2011ze} is
\beq
R^{\rm exp}_b  \equiv \frac{\Gamma(Z \to \bar{b}b)}{\Gamma(Z \to {\rm had})}=0.21629 \pm 0.00066 .
\label{Rb-exp}
\eeq
It is convenient to divide $R_b= R^{\rm SM}_b + \Delta R_b$,
where $R^{\rm SM}_b$ is predicted by the electroweak fit  as \cite{Baak:2011ze}
\beq
R^{\rm SM}_b = 0.21578 ^{+ 0.00005}_{-0.00008} \,.
\eeq
The quantity $\Delta R_b$ then encapsulates the contribution from the new physics (NP), and is represented as 
\beq
\Delta R_b = 2 R^{\rm SM}_b (1-R^{\rm SM}_b) 
{\rm Re} \left[
\frac{g^b_L \left[ \delta g^b_L\right]_{\rm NP} + g^b_R \left[ \delta g^b_R\right]_{\rm NP}}{(g^b_L)^2 + (g^b_R)^2}
\right].
\label{dRb}
\eeq
The experimental data constrains its value as
\beq
\Delta R_b = 0.00051 \pm 0.00066\,.
\label{Rb-NPconstraint}
\eeq
Eq.(\ref{dRb}) is derived straightforwardly from \cite{Hollik:1988ii} and $g^b_{L,R}$ is the SM tree level value given by
\beq
g^b_L = -\frac{1}{2} + \frac{1}{3}\sin^2\theta_W\,,\quad
g^b_R = \frac{1}{3} \sin^2\theta_W\,,
\eeq
The NP contribution $\left[ \delta g^b_{L,R} \right]_{\rm NP}$ is
\beq
\left[ \delta g^b_{L,R} \right]_{\rm NP} &=& 
\left( \left[ g^b_{L,R} \right]^{\rm tree}_{\rm BSM} - g^b_{L,R} \right)\nonumber \\
&+&
\left( \left[ \delta g^b_{L,R} \right]^{\rm 1\,loop}_{\rm BSM} - 
\left[ \delta g^b_{L,R} \right]^{\rm 1\,loop}_{\rm SM} \right).
\label{deltagfinal}
\eeq
Given a model Lagrangian, one can therefore evaluate the tree level and one loop contributions and obtain a constraint on the model. 
%
%
As a low energy effective Lagrangian, we use a Lagrangian based on the generalized hidden local symmety (GHLS) \cite{Bando:1987ym}; see also \cite{Casalbuoni:1988xm}. This means that we consider the full symmetry group to be $G_{\textrm{glo}} \times G_{\textrm{loc}}$, where $G_{\textrm{glo}} = SU(2)_L \times SU(2)_R$ and $G_{\textrm{loc}} = SU(2)_L \times SU(2)_R$. The $G_{\textrm{glo}} \times G_{\textrm{loc}}$ breaks to the diagonal $[SU(2)_V]_{\textrm{glo}}$. This symmetry breaking pattern features nine Nambu-Goldstone bosons (NGBs), which are the basic dynamical objects of the model, and which we denote as $\tilde{\pi}^a$, $\tilde{\pi}_\sigma^a$ and $\tilde{\pi}_q^a$ where $a=1,2,3$. The electroweak gauge group is embedded into the global symmetry $G_{\textrm{glo}}$ in the usual manner, and its breaking leads to the masses of the electroweak gauge bosons via absorption of the would be NGBs $\tilde{\pi}^a$. Furthermore, the vector and axial-vector mesons, $V^a_\mu$ and $A^a_\mu$, are included as dynamical gauge bosons of the hidden symmetry $G_{\textrm{loc}}$, and their masses arise via absorption of the six would-be NGBs, $\tilde{\pi}^a_\sigma$ and $\tilde{\pi}^a_q$. 

For a detailed exposition of the model, see \cite{Fukano:2011iw}. In the unitary gauge for GHLS sector, we have 
\beq
{\cal L}_{\rm GHLS}
\!\!\!&=&\!\!\!
\tr \left[ \partial_\mu \tilde\pi \partial^\mu \tilde\pi \right] + \tilde{g}^2 f^2_\sigma \tr \left[ (\tilde{V}_\mu - {\cal V}_\mu)^2\right] 
\nonumber\\[2pt]
&&\!\!\!
+ \tilde{g}^2 \chi f^2_q \tr \left[ (\tilde{A}_\mu - {\cal A}_\mu)^2\right] 
+ \tilde{g}^2 (1+\chi) f^2_q \tr \left[ \tilde{A}_\mu^2 \right] 
\nonumber\\[2pt]
&&\!\!\!
+(f^2_\pi + \chi (1 + \chi)f^2_q) \tilde{g}^2 \tr\left[\cal{A}_\mu\right]^2
\nonumber\\[2pt]
&&\!\!\!
+
\frac{2}{i} g_{\tilde{V}\pi\pi} \tr [\tilde{V}_\mu[ \partial^\mu \tilde\pi, \tilde\pi]] 
+
g_{4\pi}\tr [[ \partial_\mu \tilde\pi, \tilde\pi][ \partial^\mu \tilde\pi, \tilde\pi]]\nonumber\\[2pt]
&&\!\!\!
+ {\cal L}^{(3)}(\tilde\pi,\tilde{V},\tilde{A})
+ \cdots\,,
\nonumber\\
\eeq
where we have written the electroweak SU(2)$_L$ and U(1)$_Y$ gauge fields $\tilde{W}^a_\mu$ and $\tilde{B}_\mu$ as $\sqrt{2}\tilde{g}\cal{A}_\mu=\cal{R}_\mu-\cal{L}_\mu$ and $\sqrt{2}\tilde{g}\cal{V}_\mu=\cal{R}_\mu+\cal{L}_\mu$ in terms of ${\cal L}_\mu = g \tilde{W}^a_\mu T^a$ and ${\cal R}_\mu = g' \tilde{B}_\mu T^3$. Furthermore, we have the couplings
\beq
g_{V\pi\pi} = \frac{(1-\chi^2)f^2_\sigma}{2\sqrt{2}f^2_\pi} \tilde{g}
\,, \quad
g_{4\pi} = \left[ \frac{1}{6} - \frac{g^2_{V\pi\pi}}{\tilde{g}^2} \frac{f^2_\pi}{f^2_\sigma} \right]\frac{1}{f^2_\pi}\,.
\label{gVpp-GHLS}
\eeq
The term ${\cal L}^{(3)}$ represents triple vertices which contribute to $Z\bar{b}b$ vertex, and whose explicit form is rather lengthy. 

In Technicolor there is no direct coupling between the SM matter field and Technicolored matter or gauge fields. To understand how new physics contributions to $Z\bar{b}b$-vertex nevertheless arise, is simple: 
First, the only NGB field which can couple to the SM fermions is $\pi$, i.e. the field absorbed by the EW gauge bosons.\footnote{The field $\pi$ is in mass basis and it mainly consists of $\tilde{\pi}$ in the gauge basis.} This results in the following Yukawa coupling:
\beq
{ \cal L}_{\Sigma \bar{f}f}
=
- \bar{\psi}_L \left[ 1 + i \frac{\sqrt{2} \tilde{\pi}}{ f_\pi} \right] \!\bpm m_t & 0 \\ 0 & m_b\epm
\psi_R
+ {\rm h.c.}\,,
\label{wbGBf}
\eeq
where $\psi = (t, b)^T$ is $SU(2)$ doublet and $\psi_{L/R} \to g_{L/R} \psi_{L/R}$ under $G_{\rm glo}$. In this paper we consider only the third family quarks, and we set the $(3,3)$-component of the CKM matrix $V^{33}_{\rm CKM}$ equal to one. The contribution from ${\cal L}_{\Sigma\bar{f}f}$ can be removed by transforming to the unitary gauge. Second, the SM fermions do not couple with the vector mesons $V,A$ in the gauge eigenbasis,
\beq
&&{\cal L}_{{\cal G}\bar{f}f}
= 
e_0 \tilde{B}_\mu \bar{\psi} \gamma^\mu \bpm 2/3 & 0 \\ 0 & -1/3 \epm \psi 
\label{Lff-minimal}
\\[1ex]
\!\!\!&&\!\!\! +
\frac{e_0}{\sqrt{2} s_\theta} \left[ \tilde{W}^+_\mu \bar{t} \gamma^\mu \frac{1-\gamma_5}{2} b + {\rm h.c.} \right]
+\nonumber \\
&&\frac{e_0}{c_\theta s_\theta} \tilde{Z}_\mu 
\bar{\psi} \gamma^\mu \left[ g_L \frac{1-\gamma_5}{2} + g_R \frac{1 + \gamma_5}{2}\right] \psi
\,,\nonumber
\eeq
where $\tilde{B}_\mu, \tilde{W}^\pm_\mu$ and $\tilde{Z}_\mu$ are SM gauge bosons in the gauge eigenbasis.  The bare couplings in Eq.(\ref{Lff-minimal}) are defined so that $e_0 = gg'/\sqrt{g^2 + g'^2}$ corresponds to the bare electric charge. 

However, the propagating physical mass eigenstates, $B_\mu$, $W^\pm_\mu$ and $Z_\mu$ for the vectors will be mixtures consisting of the states  ${ \tilde{W}_\mu, \tilde{B}_\mu,\tilde{V}_\mu, \tilde{A}_\mu}$, and
 their interactions are then essentially different from the SM case. Consequentially, the bare quantities arising in (\ref{Lff-minimal}) are rescaled when we translate from gauge eigenbasis to the mass eigenbasis. 

The evaluation of $\Delta R_b$ in this model was carried out in \cite{Fukano:2011iw}. The parameters of the model are the self-coupling $\tilde{g}$ of the vector mesons, the decay constants, $f_\pi$, $f_\sigma$, and $f_q$ of the NGBs, and a dimensionless parameter $\chi$ which can be related to the oblique $S$-parameter as
\beq
S = \frac{8\pi (1-\chi^2)}{\tilde{g}^2} \,.
\label{S-WSR}
\eeq

The three decay constants are nontrivially related via the requirement that the electroweak scale has its observed value, $f_\pi\simeq 174$ GeV, and via the Weinberg sum rule, $f^2_\sigma \simeq f^2_\pi  + \chi^2 f^2_q$. These reduce the independent parameters of the model to be $\tilde{g}$, $\chi$ and one of the decay constants, which we choose to be $f_q$, and trade with $M_A$. 
Vector meson masses are given by $M^2_V \simeq \tilde{M}^2_V = \tilde{g}^2 f^2_\sigma$ and $M^2_A \simeq \tilde{M}^2_A = \tilde{g}^2 f^2_q$ at the leading order in $g/\tilde{g},\,g'/\tilde{g}$. 
In our model, the limit $\tilde{g}\rightarrow\infty$ and $f_\sigma=f_q=0$ corresponds with the Standard Model, and
in this limit we find results  consistent with the one loop results in e.g. \cite{Bamert:1996px}. Then, at finite values of the model parameters, we obtain the NP contribution to $\Delta R_b$ numerically and evaluate the contributions defined in Eq. (\ref{deltagfinal}).

\begin{figure}[htbp]
\begin{center}
\includegraphics[scale=0.8]{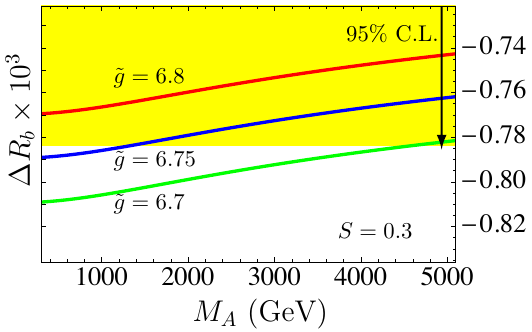}
\caption{ 
$\Delta R_b$ as a function of $M_A$ for $S=0.3$ with $\tilde{g}=6.7,6.75,6.8$. The shaded regions are $95 \% \cl$ allowed region from the constraint in Eq. (\ref{Rb-NPconstraint}). For $\tilde{g}=6.75$ the allowed region of $M_A$ is $M_A \geq 1368 \GeV$ corresponding to $M_V \geq 1497 \GeV$.
\label{Rb-MA-constraintsS03}}
\end{center}
\end{figure}

\begin{figure}[htbp]
\begin{center}
\includegraphics[scale=0.8]{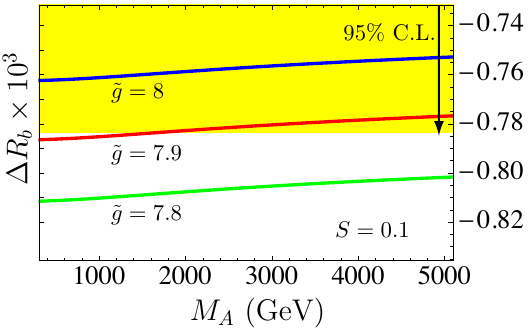}
\caption{ 
$\Delta R_b$ as a function of $M_A$ for $S=0.1$ with $\tilde{g}=7.8,7.9,8$. The shaded regions are $95 \% \cl$ allowed region from the constraint in Eq. (\ref{Rb-NPconstraint}). For $\tilde{g}=7.9$ the allowed region of $M_A$ is $M_A \geq 1670 \GeV$ corresponding to $M_V \geq 2 \TeV$.
\label{Rb-MA-constraintsS01}}
\end{center}
\end{figure}

The results for $\Delta R_b$ as a function of $M_A$ for several values of $\tilde{g}$ are shown in Figs. \ref{Rb-MA-constraintsS03} and \ref{Rb-MA-constraintsS01}, where  $S=0.3$ and $S=0.1$, respectively. 
The shaded bands in Fig. \ref{Rb-MA-constraintsS03} and \ref{Rb-MA-constraintsS01} correspond to the $95 \% \cl$ allowed region with respect to the experimental result on $\Delta R_b$. 

Of course the value of $S$ should be determined from the underlying theory. The global symmetry we have considered here corresponds to e.g. the next to minimal walking TC model \cite{Belyaev:2008yj} where one can estimate $S\sim 1/\pi$. From Figs. \ref{Rb-MA-constraintsS03} and \ref{Rb-MA-constraintsS01} we observe that  the quantitative effect is that for fixed value of $M_A$, decreasing $S$ requires stronger coupling $\tilde{g}$ in order to be compatible with the data. Alternatively, at fixed value of $\tilde{g}$, increasing $S$ allows one to saturate the constraint by lighter $M_A$. Note that throughout this paper we neglect $m_b$ contribution to the 1-loop calculations due to $m^2_b/M^2 \ll m^2_t/M^2$. Consequentially, the result for $\Delta R_b$ remains the same even if we add the neutral higgs boson to our effective Lagrangian as e.g. in \cite{Belyaev:2008yj}. 

The  contributions to $\Delta R_b$ from the vector mesons arise as $g,g'$-contribution, and since the relevant energy scale is $M_A\sim M_V\sim \Lambda_{\textrm{TC}}$, this implies that we can expect  the magnitude of $\Delta R_b$ to be of the same order as the contribution from the SM electroweak sector. 

To illustrate further the effects on phenomenology, we next consider the perturbative unitarity for scattering of longitudinal $W$-bosons under the present $R_b$ constraint.  
In the effective theory this means the $\pi\pi \to \pi \pi$ scattering  amplitude ${\cal A}_{\pi\pi \to \pi \pi}$, and for this purpose, we concentrate on $\tilde{V}\tilde{\pi}\tilde{\pi}$ and $\tilde{\pi}^4$ interaction terms in (\ref{gVpp-GHLS}), whose contribution to ${\cal A}_{\pi\pi \to \pi \pi}$ at tree level is

\beq
{\cal A}_{\pi\pi \to \pi \pi}(s,t,u)
=
3g_{4\pi} s - g^2_{V\pi\pi} \left[ \frac{u-s}{\tilde{M}^2_V - t} + \frac{t-s}{\tilde{M}^2_V - u}\right]
\,,
\nonumber
\label{amplitude}
\eeq
where $s,t,u$ are the usual Mandelstam variables $t = (-1/2)s (1 + \cos \theta)$, $u = (-1/2) s (1 - \cos \theta)$ and $\theta$ is the scattering angle. To study unitarity, the amplitude ${\cal{A}}$ is expanded in its isospin and spin components $a^I_J$. The $s$-wave, $a^0_0(s)$, given by   
\beq
a^0_0(s) &=& 
\frac{1}{64\pi}\int^1_{-1}\!\!\!\!d(\cos \theta)\left[ 3{\cal A}_{\pi\pi \to \pi \pi}(s,t,u) \right.\nonumber\\ 
&+&\left. {\cal A}_{\pi\pi \to \pi \pi}(t,s,u) +{\cal A}_{\pi\pi \to \pi \pi}(u,t,s) \right]\,,\label{00-amplitude}
\eeq
has the worst high energy behavior. For perturbative unitarity $|a^0_0| < 1/2$ should be satisfied. 
\begin{figure}[htbp]
\begin{center}
\includegraphics[scale=0.8]{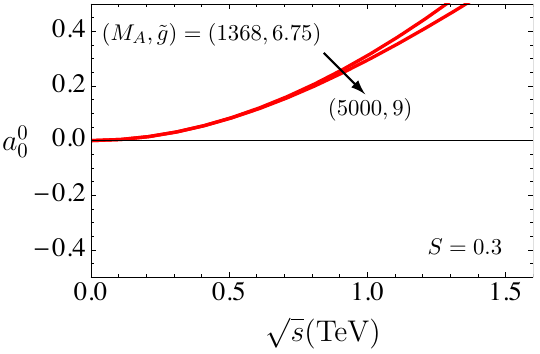} 
\caption{ 
$a^0_0$ dependence of $\sqrt{s}$ with $S=0.3$ and $95 \% \cl$ allowed region on $(M_A,\tilde{g})$-plane from $R_b$ constraint.
The lines correspond to $\tilde{g}=6.75,9$ with $M_A (\GeV)= 1368,5000$ corresponding to the $95 \% \cl$ allowed values. 
\label{Unitarity-with-RbS03}}
\end{center}
\end{figure}
\begin{figure}[htbp]
\begin{center}
\includegraphics[scale=0.8]{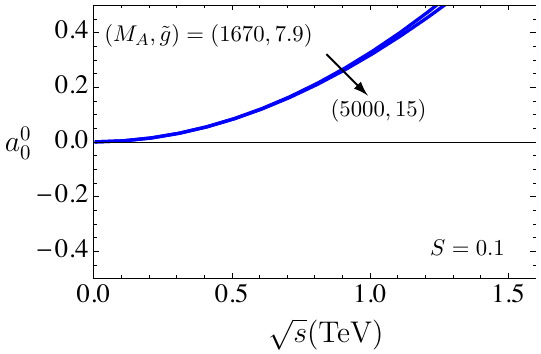} 
\caption{ 
$a^0_0$ dependence of $\sqrt{s}$ with $S=0.1$ and $95 \% \cl$ allowed region on $(M_A,\tilde{g})$-plane from $R_b$ constraint.
The lines correspond to $\tilde{g}=7.9,15$ with $M_A (\GeV)= 1670,5000$ corresponding to the $95 \% \cl$ allowed values. 
\label{Unitarity-with-RbS01}}
\end{center}
\end{figure}

In Figs. \ref{Unitarity-with-RbS03} and \ref{Unitarity-with-RbS01}, we show $a^0_0(s)$ as a function of $\sqrt{s}$ for $S=0.3$ and 0.1, respectively.  We have 
applied same relations between the model parameters as in the evaluation of $\Delta R_b$ (see Eq. (\ref{S-WSR}) and the discussion directly below it). The curves in Fig. \ref{Unitarity-with-RbS03} correspond to $(M_A,\tilde{g})=(1368,6.75)$ and $(5000,9)$ from left to right, and in Fig. \ref{Unitarity-with-RbS01} to $(M_a,\tilde{g})=(1670,7.9)$ and $(5000,15)$ similarly. The lower values were chosen on the basis of the constraint from Figs. \ref{Rb-MA-constraintsS03} and \ref{Rb-MA-constraintsS01}. Note that the dependence on $M_A$ and $\tilde{g}$ is very weak.  

Fig. \ref{Unitarity-with-RbS03} and \ref{Unitarity-with-RbS01} indicate that the tree level unitarity for $\pi \pi \to \pi \pi$ process will be broken at $\sqrt{s} \simeq 1.3 \TeV$ under the present $R_b$ constraint. We remark that this result does not change even if $g_{4\pi} = 0$,  corresponding to the cancellation of the linear growth with $s$ in (\ref{amplitude}). For alternative analysis reaching similar conclusions, see e.g. \cite{Foadi:2008xj}.


Of course, the unitarity will be protected farther out in $\sqrt{s}$ if we consider the contribution to the $\pi\pi$ scattering process from higgs boson emerging e.g. from other dynamics for the electroweak symmetry breaking like the top-quark seesaw model \cite{He:2001fz}. 
As already emphasized, the constraint from $\Delta R_b$ is insensitive to the addition of an SM Higgs -like scalar field $h$ under the approximation $m_b/m_t\ll 1$. Hence, we consider a further contribution to the low energy effective Lagrangian as \cite{Delgado:2010bb}
\beq
{\cal L}_{h \Sigma \Sigma}
=
\frac{2g_{h\pi\pi}}{m_h} \cdot h \cdot \tr \left| D_\mu \tilde{\pi} \right|^2\,.
\eeq
The above term gives the $h-\pi-\pi$ vertex and contribute to ${\cal A}_{\pi\pi \to \pi\pi}$ as  \cite{Foadi:2008xj}
\beq
\Delta_h {\cal A}_{\pi\pi \to \pi\pi}(s)
=
\frac{g^2_{h\pi\pi}}{m^2_h} \frac{s^2}{m^2_h - s}\,.
\eeq
\begin{figure}[htbp]
\begin{center}
\includegraphics[scale=0.8]{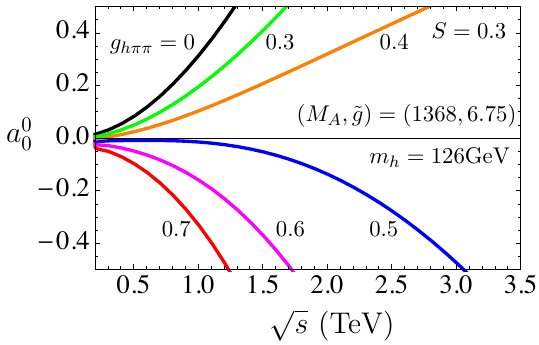}
\caption{ 
Dependence of $a^0_0$ on $\sqrt{s}$ for $m_h = 126 \GeV$ and $g_{h\pi\pi} = 0$ (black), $0.3$ (green), $0.4$ (orange), $0.5$ (blue), $0.6$ (magenta), $0.7$ (red) from left-top curves to left-bottom curves in a clockwise fashion with $S=0.3$ and $(M_A,\tilde{g}) = (1368\GeV,6.75)$. The curve with $g_{h\pi\pi} = 0$ corresponds to $(M_A,\tilde{g}) = (1368\GeV,6.75)$ curve in Fig.\ref{Unitarity-with-RbS03}.
\label{Unitarity-S03-with-higgs}}
\end{center}
\end{figure}
\begin{figure}[htbp]
\begin{center}
\includegraphics[scale=0.8]{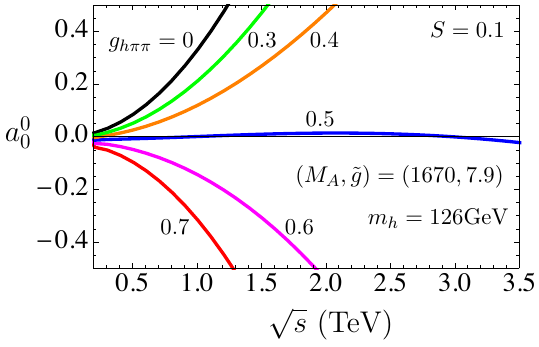} 
\caption{ 
Dependence of $a^0_0$ on $\sqrt{s}$ for $m_h = 126 \GeV$ and $g_{h\pi\pi} = 0$ (black), $0.3$ (green), $0.4$ (orange), $0.5$ (blue), $0.6$ (magenta), $0.7$ (red) from left-top curves to left-bottom curves in a clockwise fashion with $S=0.1$ and $(M_A,\tilde{g}) = (1670\GeV,7.9)$. The curve with $g_{h\pi\pi} = 0$ corresponds to $(M_A,\tilde{g}) = (1670\GeV,7.9)$ curve in Fig.\ref{Unitarity-with-RbS01}.
\label{Unitarity-S01-with-higgs}}
\end{center}
\end{figure}

In Figs. \ref{Unitarity-S03-with-higgs} and \ref{Unitarity-S01-with-higgs} we show the $\sqrt{s}$  dependence of $a^0_0$ based on interactions given by 
${\cal L_{\textrm{GHLS}}}+ {\cal L}_{h\Sigma\Sigma}$. We consider the value $m_h=126$ GeV of the higgs mass \cite{:2012gk,:2012gu}, and several values of $g_{h\pi\pi}$. 
In the figures, we again set $S=0.3$ and $S=0.1$. The values $(M_A,\tilde{g}) = (1368\GeV,6.75)$ in the case $S=0.3$ and $(M_A,\tilde{g}) = (1670\GeV,7.9)$ in the case $S=0.1$ were chosen to be compatible with the constraint from $R_b$; recall that the dependence of $a_0^0$ on $M_A$ and $\tilde{g}$ is very weak in any case.
It is easy to see from Fig. \ref{Unitarity-S03-with-higgs} and \ref{Unitarity-S01-with-higgs} that the unitarity will be protected until $\sim 3 \TeV$ and above for  $0.4 \le g_{h\pi\pi}\le 0.5$ in the case of $m_h=126$ GeV.
Based on these results we can envision the plausible scenario 
at LHC:  there exists a light Higgs with mass 126 GeV according to the observation, while the lightest vector states have masses well above 500 GeV possibly near or above one TeV. We note that strong dynamics provides a natural framework to explain the situation observed at LHC. First, the spectrum of strongly interacting composite states has the natural scale of ${\cal{O}}({\textrm{TeV}})$ and remains so far undetected. Second, near conformal strong dynamics naturally leads to a scalar state parametrically lighter than the rest of the spectrum.


We have considered a generic effective Lagrangian for dynamical electroweak symmetry breaking in a model where the global 
symmetry SU(2)$_L\times$SU(2)$_R$ breaks to SU(2)$_V$. We have shown that constraints on $R_b$ are nontrivial even in the 
absence of extended dynamics towards the generation of SM fermion masses. Of course in 
a more complete theory the contributions from the extended sectors need to be considered, but the new contributions we have 
considered in this paper will nevertheless be there and must be taken into account.
 A possible underlying model which would realize our results is  a walking Technicolor theory
with SU(3) gauge group and two sextet fermions.; this theory has naive perturbative $S$ parameter equal to 
0.3 \cite{Sannino:2004qp}. Finally we considered
perturbative unitarity of $WW$-scattering in light of the new $R_b$ constraint and demonstrated that if the Higgs was very heavy, say
of the order of 1 TeV, the unitarity would be protected only up to $\sim 1.3$ TeV for the range of parameters compatible with the $R_b$ constraint. 
Including a light Higgs, $m_h=126$ GeV, allows unitarity protection until $\sim 3$ TeV and even beyond
for suitable values of the parameters still maintaining the compatibility with $R_b$ and oblique corrections.

Our study was carried out for the GHLS type non-linear sigma model Lagrangian with a minimal coupling to SM flavors. Therefore, our results can be directly applied to several models sharing the same global symmetry at low energies. 

\end{document}